\documentclass[conference]{./lib/IEEEtran}

\ifCLASSINFOpdf
  \usepackage[pdftex]{graphicx}
\else
\fi
\usepackage[cmex10]{amsmath}
\usepackage{amssymb}
\usepackage{mathtools}
\usepackage{array}

\usepackage{algpseudocode,algorithm,algorithmicx}
\usepackage{float}
\usepackage{tabularx}
\usepackage{mathrsfs} 
\usepackage{psfrag}
\usepackage{tikz}
\usetikzlibrary{chains, arrows,shapes,positioning,arrows,decorations.markings,calc}
\usepackage{pgfplots}
\pgfplotsset{compat=newest}
\usepgfplotslibrary{groupplots}
\usepackage{gensymb}
\usepackage{xcolor}
\usepackage[shortcuts,acronym]{glossaries}
\makeglossaries % generates the acronym list
\usepackage{fixmath}
\usepackage{amssymb}
\usepackage{bm}
\usepackage{bbm}
\renewcommand{\arraystretch}{2}
\usepackage{pbox}
\usepackage{algorithm}
\usepackage{lettrine}

\renewcommand{\arraystretch}{2}	
\usepackage{url}
\usepackage{multirow}
\usepackage{graphicx}
\ifCLASSOPTIONcompsoc
\usepackage[caption=false, font=normalsize, labelfont=sf, textfont=sf]{subfig}
\else
\usepackage[caption=false, font=footnotesize]{subfig}
\fi
\usepackage{tabularx}
\pgfplotsset{compat=1.11,
	/pgfplots/ybar legend/.style={
		/pgfplots/legend image code/.code={%
			\draw[##1,/tikz/.cd,bar width=3pt,yshift=-0.2em,bar shift=0pt]
			plot coordinates {(0cm,0.8em)};},
	},}

\algnewcommand{\IIf}[1]{\State\algorithmicif\ #1\ \algorithmicthen}
\algnewcommand{\EndIIf}{\unskip\ \algorithmicend\ \algorithmicif}

\begin{document}
%
% paper title
% Titles are generally capitalized except for words such as a, an, and, as,
% at, but, by, for, in, nor, of, on, or, the, to and up, which are usually
% not capitalized unless they are the first or last word of the title.
% Linebreaks \\ can be used within to get better formatting as desired.
% Do not put math or special symbols in the title.
%\title{Advanced MEC-Aware Cell Association Strategies in 5G Mobile Systems}
%\title{MEC-based latency evaluations for V2X Cellular Communications}
\title{MEC-assisted End-to-End Latency Evaluations for C-V2X Communications}
\author{\IEEEauthorblockN{Mustafa Emara, Miltiades C. Filippou, Dario Sabella}
\IEEEauthorblockA{Next Generation and Standards, Intel Deutschland GmbH, Neubiberg, Germany \\
Email:$\{$mustafa.emara, miltiadis.filippou, dario.sabella$\}$@intel.com}
}
% use for special paper notices
%\IEEEspecialpapernotice{(Invited Paper)}
% make the title areag
\maketitle
% As a general rule, do not put math, special symbols or citations

%A
%B
\newacronym{BH}{BH}{Backhaul}
%C
\newacronym{CAM}{CAM}{Cooperative Awareness Messages}
\newacronym{CN}{CN}{Core Network}
\newacronym{C-V2X}{C-V2X}{Cellular V2X}
%D
\newacronym{DL}{DL}{Downlink}
\newacronym{DSRC}{DSRC}{Dedicated Short Range Communications}
%E
\newacronym{eNB}{eNB}{Evolved NodeB}
\newacronym{E-PDB}{E-PDB}{Extended-Packet Delay Budget}
\newacronym{E2E}{E2E}{End-to-End}
%F
\newacronym{5G}{5G}{fifth generation}
%G
%H
%I
%J
%K
%L
\newacronym{LTE}{LTE}{Long Term Evolution}
%M
\newacronym{MEC}{MEC}{Multi-access Edge Computing}
\newacronym{MBSFN}{MBSFN}{Multimedia-Broadcast Single-Frequency Networks}
%N
%O
%p
\newacronym{PPP}{PPP}{Poisson Point Process}
\newacronym{PRBs}{PRBs}{Physical Resource Blocks}
%Q
\newacronym{QoS}{QoS}{Quality-of-Service}
%R
\newacronym{RSU}{RSU}{Road Side Unit}
%S
\newacronym{SDR}{SDR}{Software Defined Radio}
\newacronym{SNR}{SNR}{Signal-to-Noise Ratio}
%T
\newacronym{TN}{TN}{Transport Network}
\newacronym{TTI}{TTI}{Transmission Time Interval}

%U
\newacronym{UL}{UL}{Uplink}
%V
\newacronym{V2I}{V2I}{Vehicle-to-Infrastructure}
\newacronym{V2P}{V2P}{Vehicle-to-Pedestrian}
\newacronym{V2V}{V2V}{Vehicle-to-Vehicle}
\newacronym{V2X}{V2X}{Vehicle-to-Everything}
\newacronym{VRU}{VRU}{Vulnerable Road User}
\newacronym{VRUs}{VRUs}{Vulnerable Road Users}
\newacronym{V2N}{V2N}{Vehicle-to-Network}
%w
%X
%Y
%Z
% !TEX root =../integration.tex
%%%%%%%%%%%%%%%%%%%%%%%%%%%%%%%%%%%%%%%%%%%%%%%%%%%%%%%%%%%%%%%%%%%%%%%%%%%%%%%%%%
% As a general rule, do not put math, special symbols or citations
% in the abstract or keywords.
\begin{abstract}
The efficient design of fifth generation (5G) mobile networks is driven by the need to support the dynamic proliferation of several vertical market segments. Considering the automotive sector, different Cellular Vehicle-to-Everything (C-V2X) use cases have been identified by the industrial and research world, referring to infotainment, automated driving and road safety. A common characteristic of these use cases is the need to exploit collective awareness of the road environment towards satisfying performance requirements. One of these requirements is the End-to-End (E2E) latency when, for instance, \ac{VRUs} inform vehicles about their status (e.g., location) and activity, assisted by the cellular network. In this paper, focusing on a freeway-based VRU scenario, we argue that, in contrast to conventional, remote cloud-based cellular architecture, the deployment of Multi-access Edge Computing (MEC) infrastructure can substantially prune the E2E communication latency. Our argument is supported by an extensive simulation-based performance comparison between the conventional and the MEC-assisted network architecture.
\end{abstract}
%%%%%%%%%%%%%%%%%%%%%%%%%%%%%%%%%%%%%%%%%%%%%%%%%%%%%%%%%%%%%%%%%%%%%%%%%%%%%%%%%%
% Note that keywords are not normally used for peerreview papers.
% no keywords
%%%%%%%%%%%%%%%%%%%%%%%%%%%%%%%%%%%%%%%%%%%%%%%%%%%%%%%%%%%%%%%%%%%%%%%%%%%%%%%%%%
% For peer review papers, you can put extra information on the cover
% page as needed:
% \ifCLASSOPTIONpeerreview
% \begin{center} \bfseries EDICS Category: 3-BBND \end{center}
% \fi
%
% For peerreview papers, this IEEEtran command inserts a page break and
% creates the second title. It will be ignored for other modes.
%\IEEEpeerreviewmaketitle
%%%%%%%%%%%%%%%%%%%%%%%%%%%%%%%%%%%%%%%%%%%%%%%%%%%%%%%%%%%%%%%%%%%%%%%%%%%%%%%%%%
% The very first letter is a 2 line initial drop letter followed
% by the rest of the first word in caps.
% 
% form to use if the first word consists of a single letter:
% \IEEEPARstart{A}{demo} file is ....
% 
% form to use if you need the single drop letter followed by
% normal text (unknown if ever used by IEEE):
% \IEEEPARstart{A}{}demo file is ....
% 
% Some journals put the first two words in caps:
% \IEEEPARstart{T}{his demo} file is ....
% 
% Here we have the typical use of a "T" for an initial drop letter
% and "HIS" in caps to complete the first word.
% You must have at least 2 lines in the paragraph with the drop letter
% (should never be an issue)
\section{Introduction}\label{section:introduction}

%Especially road safety applications
%play a very important role in vehicular communications. Road safety applications rely on short-message broadcasting in a
%vehicle’s neighborhood to inform other vehicles in order to reduce accidents on the road.

%Such packet generation can be observed, for example, in road-safety applications, when cooperative awareness messages (CAM)

%We consider latency as time interval between data generation and successful delivery to all appropriate users within MBSFN area.

%arbitrate transmission timings. In order to make evaluations simple, we also assume no bit error occurs. 

\begin{figure}
	\begin{center}
		\includegraphics[width=\columnwidth]{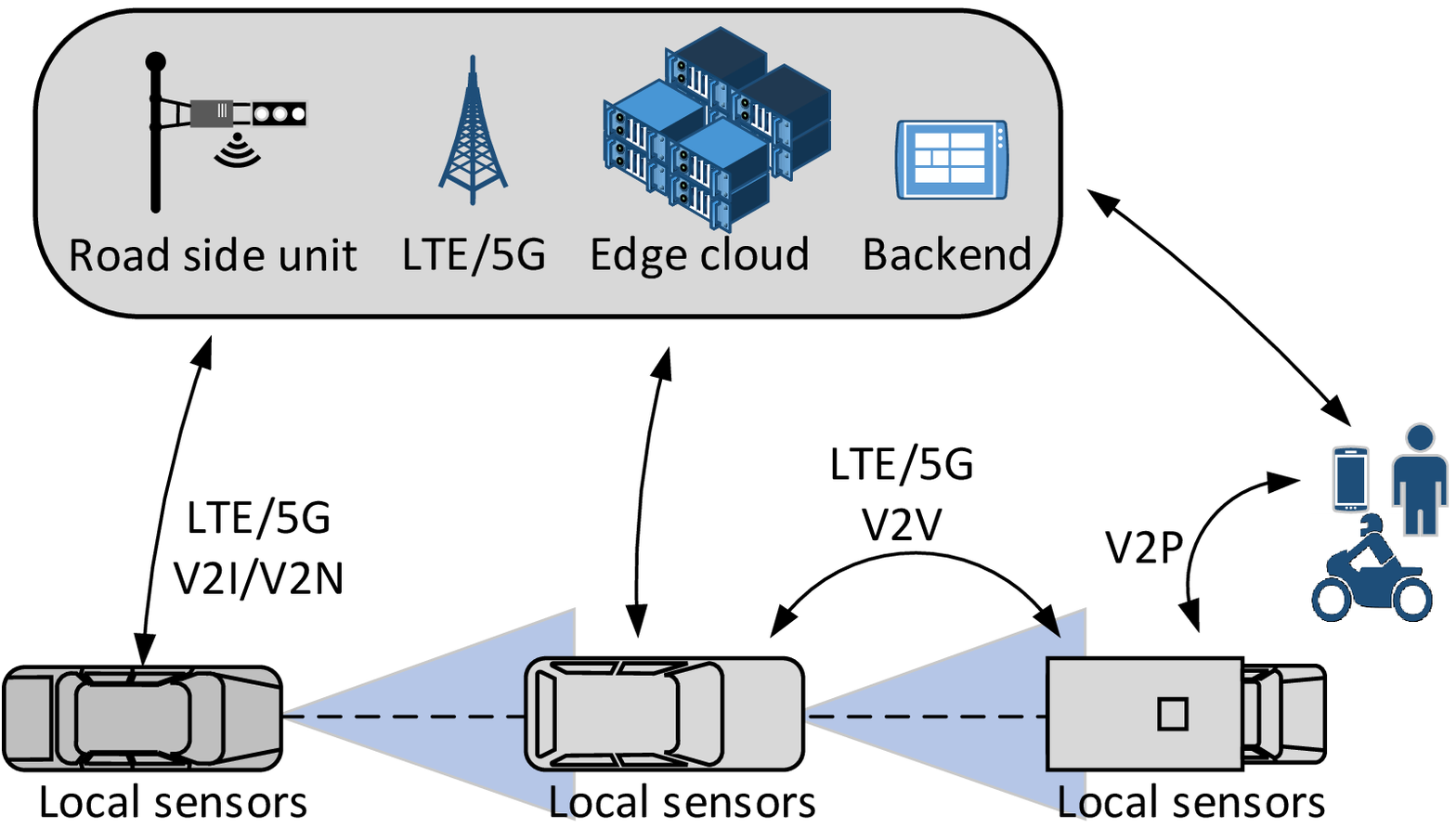}
		\caption{Envisioned \ac{5G} \ac{V2X} system.}
		\label{fig:future_systems}
	\end{center}
\end{figure}

\ac{V2X} communication paves the way for a drastically improved road safety and driving experience via reliable and low latency wireless services \cite{5GAA} \cite{2016}. The efficient \ac{V2X} system development is based on a plethora of reliably-functioning sensors, which provide an enhanced environmental perception by means of exchanging critical messages among vehicles, pedestrians and road infrastructure \cite{Gunther2015}. Such a system, as depicted in Fig.~\ref{fig:future_systems}, incorporates different information exchange paths, namely, \ac{V2I}, \ac{V2N}, \ac{V2P} and \ac{V2V} communication. These signaling paths can be either established via \ac{DSRC}, or, assisted by the cellular \ac{LTE} network providing coverage (\ac{C-V2X}), or, through an interworking of the two technologies \cite{Abboud2016}. 

Focusing on the \ac{C-V2X} technology, the architecture of the cellular network is expected to have a vital impact on the support of delay-intolerant \ac{V2X} services. This occurs, because the \ac{E2E} latency of \ac{C-V2X} signaling is limited by the quality and dimensioning of the cellular infrastructure, i.e., the capacity of backhaul connections, as well as the delays introduced by both the \ac{CN}, as well as the \ac{TN}. As one would expect, these latency bottlenecks will be more prominent for high loads corresponding to coverage areas of high vehicular/ pedestrian densities. 

To cope with such requirements, extensive research has recently taken place to enhance the advent experience of \ac{V2X} communication, with emphasis on latency shortening. For instance, in \cite{Safiulin2016}, the packet delivery latency and network utilization, focusing on an \ac{LTE} system, were investigated for \ac{MBSFN}. Furthermore, in \cite{Cattoni2015}, considering an \ac{LTE} network architecture, \ac{CN} gateway relocation is proposed for \ac{V2X} latency improvement.  Finally, with reference to implementation aspects, the authors in \cite{Lee2017} investigated latency-reduction techniques such as \ac{TTI} shortening and self-contained sub-frames in \ac{C-V2X} systems, whereas, in \cite{Cao2017}, a \ac{5G} implementation testbed for autonomous vehicles based on \ac{SDR} incorporating different solutions, was presented.

Nevertheless, in contrast to the above mentioned works, we argue that stringent latency requirements posed by the \ac{V2X} system can attract the utilization of \emph{\ac{MEC} technology}. Leveraging its ability to provide processing capabilities at the edge of the network, an overlaid \ac{MEC} deployment is expected to assist in obtaining low packet delays, due to its close proximity to end users \cite{Emara2017}. As a consequence, in this paper, concentrating on the \ac{VRU} use case, which studies the safe interaction between vehicles and non-vehicle road users (pedestrians, motorbikes, etc.) \cite{Sabella2017} via the exchange of \emph{periodic \ac{CAM}}, we aim to highlight the latency-related benefits of introducing \ac{MEC} system deployment over a state-of-the-art cellular network. Our study assumes Uu-based \ac{V2X} communication, which is one of the \ac{LTE} solutions exploiting the existing cellular infrastructure \cite{September2015}. 

The remainder of this paper is organized as follows: in Section \ref{sec:system_model}, we present an overview of the studied system model; Section \ref{sec:latency_model} provides a detailed description of the \ac{E2E} latency components and Section \ref{sec:simulation_results} presents the relevant numerical results. Finally, Section \ref{sec:Conclusion} concludes the paper. 
% !TEX root =../Integration.tex
%%%%%%%%%%%%%%%%%%%%%%%%%%%%%%%%%%%%%%%%%%%%%%%%%%%%%%%%%%%%%%%%%%%%%%%%%%%%%%%%%%
\section{System Model}\label{sec:system_model}

In this section, different aspects of the evaluation platform will be presented. First, we identify the multiple entities constituting the system setup and then we describe the \ac{VRU} use case. Finally, we review the link model and its accompanying assumptions.  

\subsection{System Setup}
\begin{figure}
	\begin{center}
		\includegraphics[width=\columnwidth]{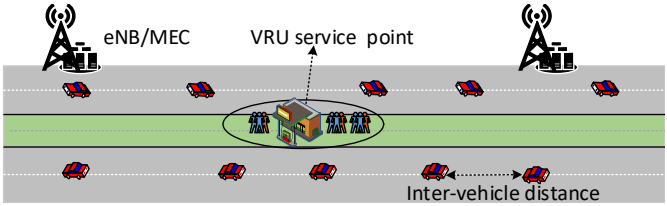}
		\caption{The investigated two-lanes freeway scenario.}
		\label{fig:full_scenario}
	\end{center}
\end{figure}

Throughout this work, a freeway road environment is assumed, consisting of one lane per direction, as shown in Fig. \ref{fig:full_scenario}. To provide a basis for possible future analytical work, which is, however, outside the scope of this paper, the vehicles are placed at the start of each system realization following a Mat\'ern hard-core point process over one dimension \cite{Haenggi2012}, with speeds drawn from a uniformly distributed random variable (i.e., $\in \mathcal{U}(\text{v}_{min}, \text{v}_{max})$). To model the inter-vehicle distance, we have resorted to the hardcore parameter of the mentioned point process, which represents the repulsion between any two generated points. Moreover, a cluster of $N$ \ac{VRUs} is on a pedestrian area between the two lanes; whereas such a populated area can be mapped to real-world scenarios like gas stations or other service points across a freeway.

At the network side, it is assumed that the focused freeway segment is under \ac{LTE} coverage; given that, for brevity, we consider a single-cell setup, the occurrence of any handover events is not taken into account by the evaluation platform. The serving \ac{eNB} is assumed to be collocated with a \ac{MEC} host of given processing capabilities, as will be explained later on.

\subsection{Vulnerable Road User - signaling model}
\begin{figure}
	\begin{center}
		\input{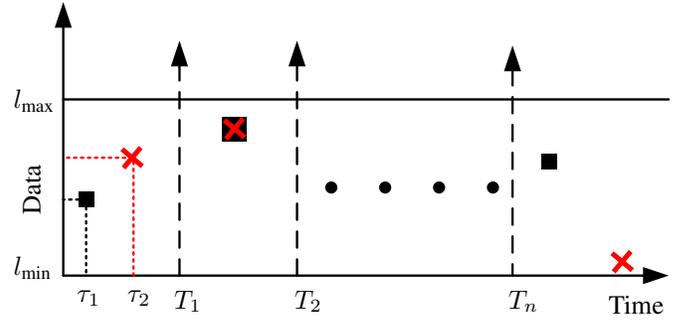}
		\caption{Packet generation procedure for two \ac{VRUs} (black square and red cross, respectively) with random transmission timing offsets.}
		\label{fig:packet_generation}
	\end{center}
\end{figure}

As highlighted in Section \ref{section:introduction}, a \ac{VRU} is assumed to interact with vehicles and, possibly, other users on the road. A straightforward example is the one of safety-related applications \cite{Kawasaki2017}, in which periodically generated \ac{VRU} messages (e.g., \ac{CAM}) can be exploited for crash prevention purposes. In order to model the generation of those periodic messages, we assume that the $k$-th \ac{VRU} generates data packets of size of $l_k \in \mathcal{U}(l_\text{min},l_\text{max})$ bits at random starting time offsets, denoted as $\tau_k$. Such \ac{CAM} transmission randomness is used to model the nature of road-safety applications. Due to the \ac{CAM} signaling periodicity, this cycle is repeated every $T$ seconds with newly generated transmission offsets. A visualization of the messaging scheme for two \ac{VRUs} is shown in Fig. \ref{fig:packet_generation}. It should be mentioned that, depending on the periodicity of packet generation and the number of \ac{VRUs} existent at the focused service point, the available \ac{UL} radio resources will need to be shared among the \ac{VRUs}.

Once a given \ac{VRU} transmits its \ac{CAM} in the \ac{UL} exploiting the Uu interface, the corresponding input packet will be processed by the \ac{MEC} host collocated with the serving \ac{eNB} and then, the processed information (output packet) will be forwarded to vehicles in the vicinity of the \ac{VRU} by means of \ac{DL} Uu-based transmission. 

According to the key results in \cite{Luoto2017}, the main challenge in designing efficient \ac{C-V2X} \ac{CAM} signaling is to serve the cell edge vehicles. Due to their low quality experienced channels, these vehicles require a large number of \ac{PRBs}, as compared to their cell-center counterparts. Therefore, accounting for the nature of \ac{CAM} messages, where the \ac{E2E} latency is dependent on the successful reception of the packets by the destined vehicles, we resort to the concept of \emph{location-based vehicle clustering}. According to this approach and, based on location availability, each \ac{VRU} defines a cluster of close-by vehicles and a cluster-based multicast transmission takes place in the \ac{DL}.  

\subsection{Link Model}
All considered vehicles and \ac{VRUs} are assumed to be served by an \ac{eNB}, based on the pathloss model adopted from the \textit{WINNER+} project \cite{September2007}, as follows
\begin{align}
\text{PL (dB)} &= 22.7\text{log}_{10}(d) - 17.3\text{log}_{10}(\tilde{h}_{\text{eNB}}) -17.3\text{log}_{10}(\tilde{h}_{\text{UE}}) \nonumber\\
&+ 2.7\text{log}_{10}(f_\text{c}) - 7.56,
\end{align}
where $d$ is the distance between the transmitter and receiver, $f_\text{c}$ is the center carrier frequency and $\tilde{h}_{\text{eNB}}$ and $\tilde{h}_{\text{VRU}}$ represent the effective antenna heights at the \ac{eNB} and \ac{VRU}, respectively. The latter quantities are computed as follows: $\tilde{h}_{\text{eNB}} = h_{\text{eNB}} - 1.0$ and  $\tilde{h}_{\text{VRU}} = h_{\text{VRU}}- 1.0$, with $h_{\text{eNB}}$ and $h_{\text{VRU}}$ being the actual antenna heights (i.e., in meters). 

Additionally, independent and identically distributed (i.i.d.) random variables are used to model the fast fading and shadowing-based attenuation phenomena. Also, it should be noted that the scheduler employed in our work equally distributes the available \ac{PRBs} over all scheduled VRUs and vehicles.

In the following section, a thorough \ac{E2E} latency analysis is presented, focusing on both the proposed, \ac{MEC}-assisted network architecture, as well as the conventional, ``distant-cloud''-based cellular architecture, which will serve as a comparison benchmark for the numerical evaluations.

% !TEX root =../Integration.tex
%%%%%%%%%%%%%%%%%%%%%%%%%%%%%%%%%%%%%%%%%%%%%%%%%%%%%%%%%%%%%%%%%%%%%%%%%%%%%%%%%%
\section{Latency Modeling}\label{sec:latency_model}
As mentioned earlier, the objective of this work is to investigate the \ac{E2E} latency performance achieved through collocated deployment of \ac{MEC} hosts and cellular network \ac{eNB}s. Towards accomplishing this aim, in this section, we model the various latency components related to \ac{CAM} transmission, routing and processing for both system approaches.

\begin{figure}
	\begin{center}
		\input{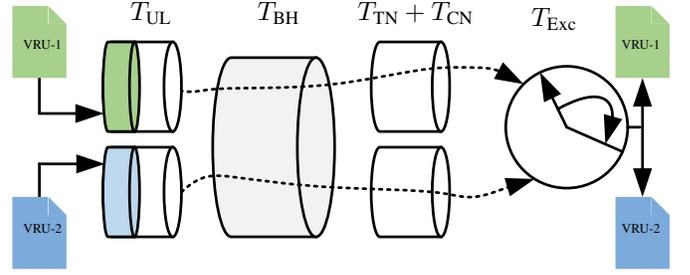}
		\caption{One-way signaling latency for two \ac{VRUs} - conventional approach.}
		\label{fig:latency_model}
	\end{center}
\end{figure}

Regarding the conventional cellular network architecture approach (Fig.~\ref{fig:latency_model}), the one-way \ac{CAM} messaging latency is modeled as $T_\text{one-way} = T_{\text{UL}} + T_{\text{BH}} + T_{\text{TN}} + T_{\text{CN}} + T_{\text{Exc}}$, where $T_{\text{UL}}$ is the radio \ac{UL} transmission latency, $T_{\text{BH}}$ is the \ac{BH} network latency, $T_{\text{TN}}$ is the \ac{TN} latency, $T_{\text{CN}}$ is the \ac{CN} latency and $T_{\text{Exc}}$ is the \ac{CAM} processing latency. Consequently, the \ac{E2E} latency, is expressed as:
 \begin{equation}\label{eq:total_latency}
  T_\text{E2E} = T_{\text{UL}} + \underbrace{2(T_{\text{BH}} + T_{\text{TN}} + T_{\text{CN}})}_\textrm{Network latency} + T_{\text{Exc}} +  T_{\text{DL}},
 \end{equation}
where, $T_{\text{DL}}$ represents the \ac{DL} transmission latency. 

For the proposed, \ac{MEC}-enabled network approach, the network latency can be avoided via processing the \ac{CAM} packets at the \ac{MEC} host, collocated with the connected \ac{eNB}. In what follows, we provide further explanations regarding the mentioned latency components.
 
\subsection{Radio Latency}
As described in Section \ref{sec:system_model}, each \ac{VRU} generates a packet for transmission within a random offset time index. The time required for the $k$-th \ac{VRU} to transmit a packet of size of $l_k$ bits to its serving \ac{eNB} is calculated as follows
\begin{align}
T_{\text{UL},k} &= \frac{l_k}{r^{UL}_k}, \\
r^{UL}_k &= \eta_k \text{log}_2(1+\text{SNR}_k),
\end{align} 
where $r^{UL}_k$ is the achievable \ac{UL} rate, $\eta_k$ is the number of allocated \ac{PRBs} and $\text{SNR}_k$ represents the received \ac{SNR} at the \ac{eNB}. Throughout this work, we assume fair resource allocation, where the total number of available \ac{PRBs}s is shared equally among the \ac{VRUs} transmitting at the same time index. As a result, the number of these \ac{VRUs}, denoted by $\hat{N}_k$, sharing the resources with the $k$-th \ac{VRU} is computed as follows
\begin{align}
\hat{N}_k &= \sum_{i=1}^{N} \mathbbm{1} (\tau_i = \tau_k), \; \forall k=\{1,2,\cdots,N \},
\end{align}
where $\mathbbm{1}(\cdot)$ is the indicator function. Due to the periodic nature of message generation, the computation of shared resources is carried out for each time window (i.e., $[T_j, T_{j+1}],\; \forall j=\{1,2,\cdots\}$). As mentioned in Section \ref{sec:system_model}, for \ac{DL} transmissions, after successful packet processing at the server, we resort to the concept of cluster-based multicast transmission \cite{Luoto2017}. The main idea is to select a set of existing vehicles in the system for transmission, in order to avoid large latencies caused by cell-edge vehicles, which would not be of high criticality for the \ac{VRU}, as the set of \ac{VRUs} is assumed to be located close to the cell center. Consequently, the vehicle cluster for the $k$-th \ac{VRU} denoted as $\mathcal{S}_k$, will consist of the $M$ closest vehicles to that \ac{VRU}. Thus, the \ac{DL} latency can be expressed as follows
\begin{equation}\label{eq:furthest}
T_{\text{DL},k} = \text{max}_{(\forall i \in \mathcal{S}_k)}\{\frac{l_k}{r^{DL}_k} \},
\end{equation}
where the maximum operator is used to measure the farthest vehicle's packet reception delay in cluster $\mathcal{S_k}$. Regardless of the \ac{eNB} location, having the $k$-th \ac{VRU} position as a reference, the maximum radio \ac{DL} latency serves as a cluster-wide metric, which is aimed to be minimized. As it will be shown later, the effect of the cluster size is significant, since the available radio resources in the \ac{DL} have to be shared among all vehicles within cluster $\mathcal{S}_k$.
\subsection{Network Latency}
As mentioned earlier, the following latency components are non-existent for the \ac{MEC}-assisted \ac{CAM} signaling case, since there is no involvement of the \ac{TN} and the \ac{CN} in \ac{CAM} packet routing. 
\subsubsection{Backhaul Latency}
The \ac{BH} latency $T_{\text{BH}}$ represents the time required for packets to be routed through the \ac{BH} network, which has a finite capacity, denoted by $C_{\text{BH}}$. It is assumed that the \ac{BH} capacity is equally shared among the $\hat{N}_k$ \ac{VRUs} concurrently uploading their messages at time instant $\tau_k$. As a result, the \ac{BH} latency for the $k$-th \ac{VRU} is 
\begin{equation}
T_{\text{BH},k} = \frac{l_k\hat{N}_k}{C_{BH}}.
\end{equation}
\subsubsection{Transport and Core Latency}
In order to provide realistic modeling of the \ac{TN} and \ac{CN} latencies, we resorted to the recent results reported in \cite{Sabella}, where a proof-of-concept was implemented for an \ac{LTE} environment with commercial terminals, running a real-time adaptive video streaming service routed through a \ac{MEC} host and several \ac{eNB} agents placed at different locations, as compared to the \ac{MEC} host position. Inspired by the results presented in this work, the two latency components are assumed to be uniformly distributed, over a range of realistic values, as it will be shown in the numerical evaluation section.
\subsection{Execution Latency}
Finally, we model the time required for processing a packet of size of $l_k$ bits at a server, either collocated with the \ac{eNB} or at the distant cloud. Assuming that the input packet requires $\beta_k$ cycles/bit for processing and the server has a processing capacity denoted by $F$, the execution latency for the $k$-th \ac{VRU} is expressed as 
\begin{equation}
 T_{\text{Exc},k} = \frac{\hat{N}_k l_k\beta_k}{F}.
\end{equation}

% !TEX root =../Integration.tex
%%%%%%%%%%%%%%%%%%%%%%%%%%%%%%%%%%%%%%%%%%%%%%%%%%%%%%%%%%%%%%%%%%%%%%%%%%%%%%%%%%
\section{Simulation Results}\label{sec:simulation_results}

In order to illustrate the latency improvements via \ac{MEC} deployment within cellular systems for \ac{V2X} communications, we provide different simulation scenarios via varying values of two main system parameters; namely, the vehicles and \ac{VRU}s spatial densities. Moreover, we also aim at observing the vehicles' cluster size impact on the experienced latency. For both the proposed and conventional cellular network architectures, the focused metric is the \ac{E2E} latency, as well as its individual components as explained in eq.~(\ref{eq:total_latency}). The values of all involved parameters are presented in Table \ref{Table:simulation_parameters}, unless otherwise stated.
	
\begin{table}
	\centering
	\caption{Simulation parameters}
	{\def\arraystretch{1}\tabcolsep=8pt
		\begin{tabular}{|c|c|c|}
			\hline
			\textit{Entity} & \textit{Parameter} & \textit{Value} \\ \hline
			\multirow{4}*{Vehicles} 
			& Speed (km/h) & $\sim \mathcal{U}(70,140)$   \\ 
			& Inter-veh. distance (m) &  10  \\
			& $\lambda$ (vehicle/m) & 0.01   \\
			& Cluster size  & 5   \\
			\hline
			\multirow{4}*{VRU}
			& Number of VRUs ($N$) & 100 \\                          
			& x-coordinates (m) & $\sim \mathcal{U}(1200,1800)$ \\ 
			& Tx power (dBm)  & 23   \\       
			& $l_k$ (kbits)  & $\sim \mathcal{U}(8,12)$ \\                     
			& $\beta_k$ (cycles/bit)  & $\sim \mathcal{U}(100,300)$ \\ 
			\hline
			\multirow{4}*{\ac{eNB} / \ac{MEC} host} 
			& Tx power (dBm) & 46 \\                           
			& Bandwidth (MHz) & 9  \\
			& $C_{BH}$ (Mbps) & 10  \\
			& $F$ (Gcycles/sec) & 9  \\  	
			\hline
			\multirow{10}*{General}
			& Frequency (GHz) & 5.9 \\
			& Number of lanes &  2  \\
			& Lane length (km) & 3 \\
			& Lane width (m) & 4 \\
			& Pathloss exponent & 3 \\ 
			& Shadowing std. dev. (dB) & 3 \\
			& Fast fading std. dev. (dB) & 4 \\
			& Thermal noise (dBm) & -110 \\
			& Additional losses (dB) & 15 \\
			& \ac{TN}+\ac{CN} latency (ms) & $\sim \mathcal{U}(15,35)$\\ 
			\hline
		\end{tabular}
	}
	\label{Table:simulation_parameters}
\end{table}
\subsection{Effect of VRU Density}
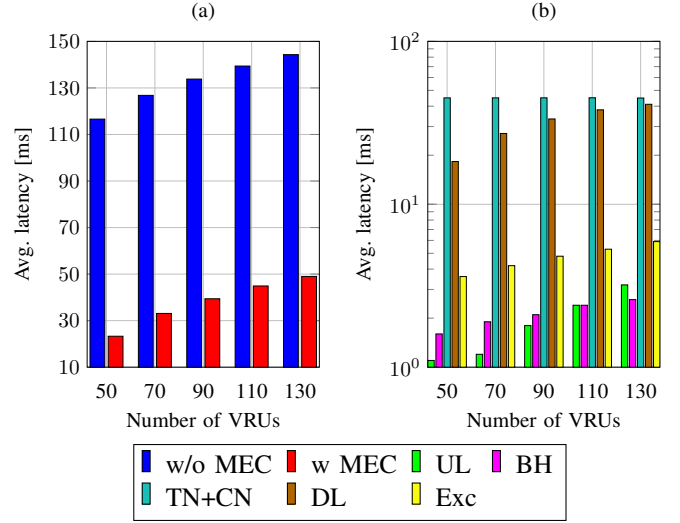
\begin{figure} 
	\centering
	% This file was created by matlab2tikz.
%
%The latest updates can be retrieved from
%  http://www.mathworks.com/matlabcentral/fileexchange/22022-matlab2tikz-matlab2tikz
%where you can also make suggestions and rate matlab2tikz.
%
\definecolor{mycolor1}{rgb}{0.24220,0.15040,0.66030}%
\definecolor{mycolor2}{rgb}{1,0.0,1.0}%
\definecolor{mycolor3}{rgb}{0.09640,0.75000,0.71204}%
\definecolor{mycolor4}{rgb}{0.69, 0.4, 0.0}%
\definecolor{mycolor5}{rgb}{0.97690,0.98390,0.08050}%

	\begin{tikzpicture}[scale=0.8]
\begin{groupplot}[group style={
	group name=myplot,
	group size= 2 by 1, horizontal sep=1.8cm},height=3cm, width= 0.8*\columnwidth]
\nextgroupplot[title={(a)}, align=left,
width  = 0.3*\textwidth,
height = 7cm,
major x tick style = transparent,
xlabel={Number of VRUs},
ylabel={Avg. latency [ms]},
ybar=4*\pgflinewidth,
bar width=7pt,
ymajorgrids = true,
xmajorgrids = true,
symbolic x coords={50,70,90,110,130},
scaled y ticks = false,
xtick={50,70,90,110,130},
ytick={10, 30, 50, 70, 90, 110, 130, 150},
legend to name=grouplegend1,
legend cell align=left,
legend style={
	legend columns=4,fill=none,draw=black,anchor=center,align=left, column sep=0.13cm},
ymin=10, 
ymax=150]

\addlegendimage{style={black,fill=blue,mark=none}}
\addlegendentry{w/o MEC}
\addlegendimage{style={black,fill=red,mark=none}}
\addlegendentry{w MEC}
\addlegendimage{style={black,fill=green,mark=none}}
\addlegendentry{UL}
\addlegendimage{style={black,fill=mycolor2,mark=none}}
\addlegendentry{BH}
\addlegendimage{style={black,fill=mycolor3,mark=none}}
\addlegendentry{TN+CN}
\addlegendimage{style={black,fill=mycolor4,mark=none}}
\addlegendentry{DL}
\addlegendimage{style={black,fill=mycolor5,mark=none}}
\addlegendentry{Exc}

\addplot[style={black,fill=blue,mark=none}] coordinates {(50,116.6) (70,126.8) (90,133.8) (110,139.4) (130,144.3)};

\addplot[style={black,fill=red,mark=none}] coordinates {(50,23.3) (70,33.1) (90,39.4) (110,44.9) (130,49)};

\coordinate (c1) at (rel axis cs:0,0);

\nextgroupplot[title={(b)}, align=left,
width  = 0.3*\textwidth,
height = 7cm,
major x tick style = transparent,
xlabel={Number of VRUs},
ylabel={Avg. latency [ms]},
ybar=2*\pgflinewidth,
bar width=3pt,
ymajorgrids = true,
xmajorgrids = true,
symbolic x coords={50,70,90,110,130},
scaled y ticks = false,
xtick={50,70,90,110,130},
ymode=log,
ymin=1,
ymax=100, 
]
\addplot[style={black,fill=green,mark=none}] coordinates    {(50,1.1) (70,1.2) (90,1.8) (110,2.4) (130,3.2)};
\addplot[style={black,fill=mycolor2,mark=none}] coordinates {(50,1.6) (70,1.9) (90,2.1) (110,2.4) (130,2.6)};
\addplot[style={black,fill=mycolor3,mark=none}] coordinates {(50,45) (70,45.01) (90,45.03) (110,45.06) (130,44.9)};
\addplot[style={black,fill=mycolor4,mark=none}] coordinates {(50,18.3) (70,27.2) (90,33.4) (110,38) (130,41.1)};
\addplot[style={black,fill=mycolor5,mark=none}] coordinates {(50,3.6) (70,4.2) (90,4.8) (110,5.3) (130,5.9)};

\coordinate (c2) at (rel axis cs:0.5,0);
\end{groupplot}

\coordinate (c3) at ($(c1)!.5!(c2)$);
\node[below] at (c3 |- current bounding box.south)
{\ref{grouplegend1}};
\end{tikzpicture}
	\caption{(a) Average \ac{E2E} latency for increasing number of VRUs. (b) Component-wise latency breakdown.}
	\label{fig:inc_VRUs} 
\end{figure}

First, we look into the case of increasing \ac{VRU}s. As explained in the previous sections, each \ac{VRU} is assigned a random timing offset for transmission. Thus, the generated periodic message traffic increases accordingly with the \ac{VRU}s.  In Figure~\ref{fig:inc_VRUs}, the average \ac{E2E} signaling latency with and without \ac{MEC} host deployment is shown both as a whole and component-wise. Clearly, \ac{MEC} utilization provides a lower \ac{E2E} latency (the observed gains are in the range of 66\%-80\%), due to the exploitation of processing resource proximity offered by the \ac{MEC} host. Additionally, we observe an increasing behavior of the latency along with the \ac{VRU} density, which is due to the increasing demand of the available resources. First, for the radio transmission latency components, as the number of \ac{VRU}s increases, the available resources per \ac{VRU} decrease, due to the equal allocation assumption. Similar explanations hold for the \ac{BH} and the execution latencies. It should be noted that the \ac{TN} and \ac{CN} latencies were modeled as random variables, independent of the system parameters, which can be further modified in future work. 

\subsection{Effect of Vehicles Density}

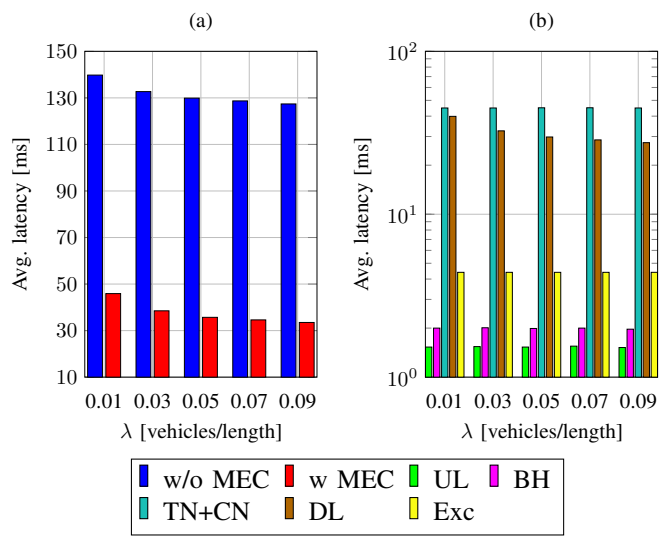
\begin{figure} 
	\centering
	% This file was created by matlab2tikz.
%
%The latest updates can be retrieved from
%  http://www.mathworks.com/matlabcentral/fileexchange/22022-matlab2tikz-matlab2tikz
%where you can also make suggestions and rate matlab2tikz.
%
\definecolor{mycolor1}{rgb}{0.24220,0.15040,0.66030}%
\definecolor{mycolor2}{rgb}{1,0.0,1.0}%
\definecolor{mycolor3}{rgb}{0.09640,0.75000,0.71204}%
\definecolor{mycolor4}{rgb}{0.69, 0.4, 0.0}%
\definecolor{mycolor5}{rgb}{0.97690,0.98390,0.08050}%

	\begin{tikzpicture}[scale=0.8]
\begin{groupplot}[group style={
	group name=myplot,
	group size= 2 by 1, horizontal sep=1.8cm},height=3cm, width= 0.8*\columnwidth]
\nextgroupplot[title={(a)}, align=left,
width  = 0.3*\textwidth,
height = 7cm,
major x tick style = transparent,
xlabel={$\lambda$ [vehicles/length]},
ylabel={Avg. latency [ms]},
ybar=4*\pgflinewidth,
bar width=7pt,
ymajorgrids = true,
xmajorgrids = true,
symbolic x coords={0.01,0.03,0.05,0.07,0.09},
scaled y ticks = false,
xtick={0.01, 0.03, 0.05, 0.07, 0.09},
ytick={10, 30, 50, 70, 90, 110, 130, 150},
legend to name=grouplegend1,
legend cell align=left,
legend style={
	legend columns=4,fill=none,draw=black,anchor=center,align=left, column sep=0.13cm},
ymin=10, 
ymax=150]

\addlegendimage{style={black,fill=blue,mark=none}}
\addlegendentry{w/o MEC}
\addlegendimage{style={black,fill=red,mark=none}}
\addlegendentry{w MEC}
\addlegendimage{style={black,fill=green,mark=none}}
\addlegendentry{UL}
\addlegendimage{style={black,fill=mycolor2,mark=none}}
\addlegendentry{BH}
\addlegendimage{style={black,fill=mycolor3,mark=none}}
\addlegendentry{TN+CN}
\addlegendimage{style={black,fill=mycolor4,mark=none}}
\addlegendentry{DL}
\addlegendimage{style={black,fill=mycolor5,mark=none}}
\addlegendentry{Exc}

\addplot[style={black,fill=blue,mark=none}] coordinates {(0.01,139.8) (0.03,132.7) (0.05,129.9) (0.07,128.7) (0.09,127.4)};

\addplot[style={black,fill=red,mark=none}] coordinates {(0.01,45.9) (0.03,38.5) (0.05,35.7) (0.07,34.6) (0.09,33.5)};

\coordinate (c1) at (rel axis cs:0,0);

\nextgroupplot[title={(b)}, align=left,
width  = 0.3*\textwidth,
height = 7cm,
major x tick style = transparent,
xlabel={$\lambda$ [vehicles/length]},
ylabel={Avg. latency [ms]},
ybar=2*\pgflinewidth,
bar width=3pt,
ymajorgrids = true,
xmajorgrids = true,
symbolic x coords={0.01,0.03,0.05,0.07,0.09},
scaled y ticks = false,
xtick={0.01, 0.03, 0.05, 0.07, 0.09},
ymode=log,
ymin=1,
ymax=100, 
]
\addplot[style={black,fill=green,mark=none}] coordinates {(0.01,1.53) (0.03,1.54) (0.05,1.53)  (0.07,1.55)  (0.09,1.52)};
\addplot[style={black,fill=mycolor2,mark=none}] coordinates {(0.01,2)    (0.03,2.01) (0.05,1.99)  (0.07,2)     (0.09,1.97)};
\addplot[style={black,fill=mycolor3,mark=none}] coordinates {(0.01,44.9) (0.03,44.9) (0.05,45.02) (0.07,45.04) (0.09,44.9)};
\addplot[style={black,fill=mycolor4,mark=none}] coordinates {(0.01,39.9) (0.03,32.5) (0.05,29.8)  (0.07,28.6)  (0.09,27.5)};
\addplot[style={black,fill=mycolor5,mark=none}] coordinates {(0.01,4.4)  (0.03,4.4)  (0.05,4.4)   (0.07,4.4)   (0.09,4.4)};

\coordinate (c2) at (rel axis cs:0.5,0);
\end{groupplot}

\coordinate (c3) at ($(c1)!.5!(c2)$);
\node[below] at (c3 |- current bounding box.south)
{\ref{grouplegend1}};
\end{tikzpicture}
	\caption{(a) Average \ac{E2E} latency for increasing vehicles' deployment densities. (b) Component-wise latency breakdown.}
	\label{fig:inc_vehicles} 
\end{figure}

In this part, an alternative scenario of fixing the number of \ac{VRU}s and increasing the spatial density of the vehicles is studied, as per Fig.~\ref{fig:inc_vehicles}. Since the \ac{VRU}s in the investigated use case are the active agents and the vehicles are the passive ones, i.e., transmission is always initiated by the \ac{VRU}s, the \ac{E2E} latency is dependent on the vehicles' spatial density. As discussed in Section \ref{sec:system_model}, the vehicles' density (i.e., $\lambda$) only plays a role in the radio \ac{DL} latency. Since a location-based multicast transmission is employed, where the cluster size (i.e., $|\mathcal{S}_k|$) is fixed, as the number of vehicles increases, the probability to have the cluster closer to the \ac{VRU} of interest will increase as well. Hence, as expected, the \ac{DL} latency decreases with increasing $\lambda$. 

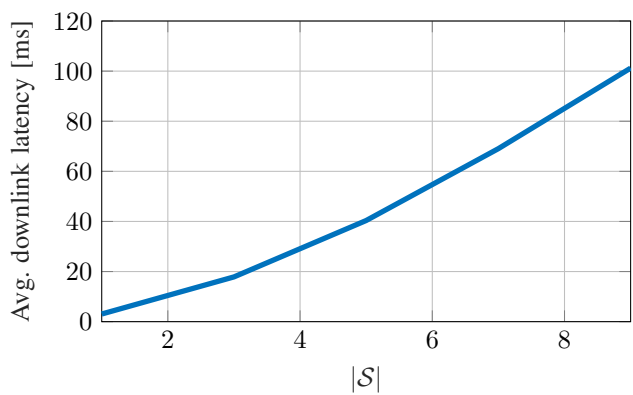
\begin{figure} 
	\centering
	\definecolor{mycolor1}{rgb}{0.00000,0.44700,0.74100}%
\begin{tikzpicture}

\begin{axis}[%
width=0.8\columnwidth,
height=4cm,
at={(0.758in,0.481in)},
scale only axis,
xmin=1,
xmax=9,
xlabel style={font=\color{white!15!black}},
xlabel={$|\mathcal{S}|$},
ymin=0,
ymax=120,
ytick={0,20,40,60,80,100,120},
ylabel style={font=\color{white!15!black}},
ylabel={Avg. downlink latency [ms]},
axis background/.style={fill=white},
xmajorgrids,
ymajorgrids
]
\addplot [color=mycolor1, line width=2.0pt, forget plot]
  table[row sep=crcr]{%
1	3.05\\
3	17.86\\
5	40.37\\
7	69.06\\
9	101.2\\
};
\end{axis}
\end{tikzpicture}%
	\caption{Average cluster-related radio \ac{DL} latency as a function of the vehicle cluster size.}
	\label{fig:clustersize} 
\end{figure}

Since the cluster size highly affects the \ac{E2E} latency through its contribution to the \ac{DL} radio latency, the experienced \ac{DL} latency for increasing vehicle cluster sizes is simulated and presented in Fig.~\ref{fig:clustersize}. Due to the definition of the \ac{DL} latency (eq. (\ref{eq:furthest})) and its dependence on the cluster's farthest vehicle to successfully receive the packet, as the cluster size increases, the probability of vehicles being far from the focused \ac{VRU} will increase as well. As a result, this explains the increasing fashion of the radio \ac{DL} latency, which is as depicted in Fig.~\ref{fig:clustersize}. 

% !TEX root =../Integration.tex
%%%%%%%%%%%%%%%%%%%%%%%%%%%%%%%%%%%%%%%%%%%%%%%%%%%%%%%%%%%%%%%%%%%%%%%%%%%%%%%%%%
\section{Conclusion}\label{sec:Conclusion}
In this paper, we have investigated the problem of improving the timeliness of collective road awareness, concentrating on the VRU use case and focusing on a freeway segment under cellular network coverage. With the aim of minimizing \ac{E2E} signaling latency, we have proposed a MEC-assisted network architecture, according to which MEC hosts are collocated with \ac{eNB}s, thus, they can receive and process VRU messages at the edge of the access network. Towards quantifying the benefits of the new approach, we have defined the latencies related to radio transmission and message processing, driven by realistic assumptions. By means of numerical evaluation, it has been observed that, for some of the investigated system parameterizations, the proposed overlaid deployment of MEC hosts offers up to 80\%  average gains in latency reduction, as compared to the conventional network architecture. It is interestingly shown that performance benefits remain significant for different vehicle/ VRU deployment densities, as well as for different vehicle cluster sizes when VRU-to-vehicle distance-dependent multi-cast signaling is performed.

\section*{Acknowledgment}
The research leading to these results has been performed under the framework of the Horizon 2020 project ONE5G (ICT-760809) receiving funds from the European Union.
% !TEX root =../quant_cell_det_jnl.tex
%%%%%%%%%%%%%%%%%%%%%%%%%%%%%%%%%%%%%%%%%%%%%%%%%%%%%%%%%%%%%%%%%%%%%%%%%%%%%%%%%%
% Can use something like this to put references on a page
% by themselves when using endfloat and the captionsoff option.
%\ifCLASSOPTIONcaptionsoff
%  \newpage
%\fi
\bibliographystyle{./lib/IEEEtran.cls}
% argument is your BibTeX string definitions and bibliography database(s)
%\bibliography{IEEEabrv,../bib/paper}
\bibliography{./literature/Literature_Local}

% Generated by IEEEtran.bst, version: 1.14 (2015/08/26)
\begin{thebibliography}{10}
\providecommand{\url}[1]{#1}
\csname url@samestyle\endcsname
\providecommand{\newblock}{\relax}
\providecommand{\bibinfo}[2]{#2}
\providecommand{\BIBentrySTDinterwordspacing}{\spaceskip=0pt\relax}
\providecommand{\BIBentryALTinterwordstretchfactor}{4}
\providecommand{\BIBentryALTinterwordspacing}{\spaceskip=\fontdimen2\font plus
\BIBentryALTinterwordstretchfactor\fontdimen3\font minus
  \fontdimen4\font\relax}
\providecommand{\BIBforeignlanguage}[2]{{%
\expandafter\ifx\csname l@#1\endcsname\relax
\typeout{** WARNING: IEEEtran.bst: No hyphenation pattern has been}%
\typeout{** loaded for the language `#1'. Using the pattern for}%
\typeout{** the default language instead.}%
\else
\language=\csname l@#1\endcsname
\fi
#2}}
\providecommand{\BIBdecl}{\relax}
\BIBdecl

\bibitem{5GAA}
``{5G Automative Association},'' \url{http://5gaa.org/}.

\bibitem{2016}
``Leading the world to {5G}: Cellular vehicle-to-everything {(C-V2X)}
  technologies,'' Qualcomm [Online]. Available:
  https://www.qualcomm.com/media/documents/files/cellular-vehicle-to-everything-c-v2x-technologies.pdf,
  Tech. Rep., 2016.

\bibitem{Gunther2015}
H.~j.~Gunther, O.~Trauer, and L.~Wolf, ``The potential of collective perception
  in vehicular ad-hoc networks,'' in \emph{2015 14th International Conference
  on ITS Telecommunications (ITST)}, Dec. 2015, pp. 1--5.

\bibitem{Abboud2016}
K.~Abboud, H.~A. Omar, and W.~Zhuang, ``Interworking of {D}{S}{R}{C} and
  cellular network technologies for {V}2{X} communications: A survey,''
  \emph{IEEE Transactions on Vehicular Technology}, vol.~65, no.~12, pp.
  9457--9470, Dec. 2016.

\bibitem{Safiulin2016}
I.~Safiulin, S.~Schwarz, T.~Philosof, and M.~Rupp, ``Latency and resource
  utilization analysis for {V}2{X} communication over {L}{T}{E} {M}{B}{S}{F}{N}
  transmission,'' in \emph{WSA 2016; 20th International ITG Workshop on Smart
  Antennas}, Mar. 2016, pp. 1--6.

\bibitem{Cattoni2015}
A.~F. Cattoni, D.~Chandramouli, C.~Sartori, R.~Stademann, and P.~Zanier,
  ``Mobile low latency services in 5{G},'' in \emph{2015 IEEE 81st Vehicular
  Technology Conference (VTC Spring)}, May 2015, pp. 1--6.

\bibitem{Lee2017}
K.~Lee, J.~Kim, Y.~Park, H.~Wang, and D.~Hong, ``Latency of cellular-based
  {V}2{X}: Perspectives on {T}{T}{I}-proportional latency and
  {T}{T}{I}-independent latency,'' \emph{IEEE Access}, vol.~5, pp.
  15\,800--15\,809, 2017.

\bibitem{Cao2017}
H.~Cao, S.~Gangakhedkar, A.~R. Ali, M.~Gharba, and J.~Eichinger, ``A testbed
  for experimenting 5{G}-{V}2{X} requiring ultra reliability and low-latency,''
  in \emph{WSA 2017; 21th International ITG Workshop on Smart Antennas}, Mar.
  2017, pp. 1--4.

\bibitem{Emara2017}
M.~{Emara}, M.~C. {Filippou}, and D.~{Sabella}, ``{MEC-aware Cell Association
  for {5G} Heterogeneous Networks},'' \emph{ArXiv e-prints}, Nov. 2017.

\bibitem{Sabella2017}
{D. Sabella et al.}, ``Toward fully connected vehicles: Edge computing for
  advanced automotive communications,'' [Online]. Available:
  http://5gaa.org/news/toward-fully-connected-vehicles-edge-computing-for-advanced-automotive-communications/,
  Tech. Rep., 2017.

\bibitem{September2015}
``3{G}{P}{P} {T}{R} 36.885 {V}14.0.0, study on {L}{T}{E}-based {V}2{X} services
  ({R}elease 14),'' [Online]. Avalible:http://www.3gpp.org/, Tech. Rep.,
  September 2015.

\bibitem{Haenggi2012}
M.~Haenggi, \emph{Stochastic Geometry for Wireless Networks}, 1st~ed.\hskip 1em
  plus 0.5em minus 0.4em\relax New York, NY, USA: Cambridge University Press,
  2012.

\bibitem{Kawasaki2017}
R.~Kawasaki, H.~Onishi, and T.~Murase, ``Performance evaluation on {V}2{X}
  communication with {P}{C}5-based and {U}u-based {L}{T}{E} in crash warning
  application,'' in \emph{2017 IEEE 6th Global Conference on Consumer
  Electronics (GCCE)}, Oct. 2017, pp. 1--2.

\bibitem{Luoto2017}
{P. Luoto et al.}, ``Vehicle clustering for improving enhanced
  {L}{T}{E}-{V}2{X} network performance,'' in \emph{2017 European Conference on
  Networks and Communications (EuCNC)}, Jun. 2017, pp. 1--5.

\bibitem{September2007}
``{WINNER II} channel models, {D}1.1.2 v1.0,'' [Online]. Available:
  https://cept.org/files/8339/winner2\%20-\%20final\%20report.pdf, Tech. Rep.

\bibitem{Sabella}
{D. Sabella et al.}, ``A hierarchical {MEC} architecture: experimenting the
  {RAVEN} use case,'' \emph{CLEEN2018 workshop, June 2018}.

\end{thebibliography}

\end{document}